\documentclass[letterpaper, 10 pt, conference]{ieeeconf}  

\IEEEoverridecommandlockouts                              

\usepackage[bookmarks=false]{hyperref}
\usepackage{amsmath}
\usepackage{amssymb}
\usepackage{kikuuwe}
\usepackage[linesnumbered,ruled,vlined]{algorithm2e}
\usepackage{arydshln}
\usepackage{graphicx}
\usepackage{eqnarraycompact}
\SetKwInOut{Parameter}{parameter}

\newcommand{\bsbeq}{\begin{subequations}}
\newcommand{\esbeq}{\end{subequations}}
\newcommand\bbR{{\mathbb{R}}}

\newif\ifshowdetail\showdetailfalse

\newcommand{\argmin}{\operatornamewithlimits{argmin}}

\newcommand\proj{\mathrm{proj}}

\allowdisplaybreaks
\newtheorem{lemma}{Lemma}

\newtheorem{rem}{Remark}

\IEEEoverridecommandlockouts                              
\title{\LARGE \bf
Chattering-Free Implementation of Continuous Terminal Algorithm with Implicit Euler Method
}

\author{Xiaogang Xiong$^{1}$, Wei Chen$^{2}$ and Guohua Jiao$^{2}$, Shanhai Jin$^{3}$, and Shyam Kamal$^{4}$
\thanks{$^{1}$ X. Xiong is with Harbin Institute of Technology,
       Shenzhen, P.R., China
        {\tt\small heha@foxmail.com}}%
\thanks{$^{2}$ W. Chen and G. Jiao with Shenzhen Institutes of Advanced Technology,Chinese Academy of Sciences, Shenzhen, P.R., China}
\thanks{$^{3}$S. Jin is with School of Engineering, Yanbian University, P.R., China}
\thanks{$^{4}$S. Kamal is with IIT(BHU), India}
}

\begin{document}

\maketitle
\thispagestyle{empty}
\pagestyle{empty}

\begin{abstract}
This paper proposes an efficient implementation for a continuous terminal algorithm (CTA). Although CTA is a continuous version of the famous twisting algorithm (TA), the conventional implementations of this CTA still suffer from chattering, especially when the gains and the time-step sizes are selected very large. The proposed implementation is based on an implicit Euler method and it totally suppresses the chattering. The proposed implementation is compared with the conventional explicit Euler implementation through simulations. It shows that the proposed implementation is very efficient and the chattering is suppressed both in the control input and output.

\end{abstract}

\section{Introduction} \label{sec:Cont-CTSMC}
Terminal Sliding mode control (TSMC) has been recognized as one of potentially useful control schemes due to
its finite-time convergence, tracking accuracy, robustness against parametric uncertainty and external disturbances and simple tuning of parameters \cite{Davila_2005_Oberver,Levant_1993_Order,Levant_2007_principle}. In practice, the main drawback of TSMC is numerical chattering which could cause damages to the actuators of systems and deteriorate the control performance. Several solutions have been proposed to alleviate the numerical chattering, such as higher order sliding mode (HOSM) \cite{Levant_2007_principle,Shyam_2015_Continuous}, adaptive sliding mode designs \cite{Xiong_2018_Adaptive01,Xiong_2018_Adaptive02,Taleb_2014_HOSMC,Utkin_2013_ASTA}, and implicit Euler methods \cite{Acary_2012_Euler,ACARY2010284}.

Recently, some kinds of continuous terminal algorithms (CTAs) have been proposed \cite{Shyam_2016_CTSMC,CTA_2015_conference,CTA_2017_journal} by using the homogeneity properties. Their structure are very similar to the famous super-twisting algorithms (STA). However, the relative degrees of the CTAs is higher than STA. The control input becomes smoother because the discontinuous signum functions are integrated, and this is the reason why they are called ``continuous" terminal algorism. However, the implementation of CTAs with traditional explicit Euler methods still suffer from numerical chattering, as shown the Example section of \cite{Shyam_2016_CTSMC}. On the other hand, sliding-mode control systems have been associated with differential inclusions. This paper tries to propose discrete-time implementation algorithms of CTAs based on the differential inclusions.

The proposed algorithms are based on the implicit-Euler discretization. In contrast to explicit-Euler implementations, which are prone to chattering, the implicit-Euler implementations do not result in chattering \cite{Shyam_2015_Continuous,Utkin_2013_ASTA,Huber_2016_Implicit}. Huber et al proposed implicit-Euler implementation of twisting control \cite{Huber_2016_Implicit} based on ZOH (Zero-Order Hold) discretization and AVI (Affine Variational Inequality) solver. The twisting control is one of HOSM with simple structure. Xiong et al proposed a simpler calculation for the twisting control based on the implicit Euler discretization. However, as far as the authors are aware, there is few work that proposes implicit Euler discretization of the other HOSM algorithms such as the CTAs proposed in \cite{Shyam_2016_CTSMC,CTA_2015_conference,CTA_2017_journal}. The difficulty lies in that the implicit Euler dicretization results in a complicate nonlinear implicit functions and stability analysis. This paper tries to explore the implicit Euler dicretizations of CTAs in a chattering-free manner while keeping its robustness.

This paper is organized as follows. Section~\ref{sec:math} introduces some mathematical preliminaries that will be used in this paper. New discrete-time algorithms of the continuous terminal algorithm are proposed in Section~\ref{sec:control}, respectively. The effectiveness of the proposed algorithms are illustrated in Section~\ref{sec:simulation} through simulation results. Some concluding remarks are given in Section~\ref{sec:conclusion}.

\section{Mathematical Preliminaries}
\label{sec:math}
Let $\calA$ be a closed interval of real numbers. This paper uses the following function:
\beqa
\proj_{\calA}(x)&\define& \argmin_{\xi\in\calA}(\xi-x)^2.
\eeqa
This function $\proj_{\calA}$ can be referred to as the projection onto the set $\calA$. If the set $\calA$ is written as $\calA=[A,B]$ where $A\leq B$, this function can be written as follows:
\beqa
\proj_{[A,B]}(x)&=& \lb\{\barr{ll} B & \mbox{if }x >B  \\ x &\mbox{if } x\in[A,B] \\ A &\mbox{if } x<A. \earr\rb.
\eeqa
This paper also uses the following definition of signum function:
\beqa \dfeq{sgn}
\sgn(x)&\define&\lb\{\barr{ll} [-1,1]\quad&\mbox{if }x=0 \\ x/|x| \quad &\mbox{if }x\neq 0. \earr\rb.
\eeqa
 This kind of set-valued definition has been employed by many previous papers \cite{Acary_2012_Euler,ACARY2010284,Brogliato_2018_STA}. With a non-negative scalar $F\geq 0$, the following relation is satisfied:
\beqa \dfeq{sgnsat}
x\in F\sgn(y-x)  \iff x = \proj_{[-F,F]}(y),
\eeqa
of which proofs are found in previous papers \cite{Acary_2012_Euler,Kikuuwe_2010_PSMC}.
\par

The remainder of this paper employs the following lemmas:
\begin{lemma}\label{lm:lmx}
With $A>B>0$, let us define $\calC\define [A-B,A+B]$. Then, for all $x,y\in\bbR$, the following two statements are equivalent to each other:
\beqa
y \in [-A,A] + B\sgn(x-y)  \dfeq{stt1a}\\
y \in [\proj(-\calC,x),\proj(\calC,x)]. \dfeq{stt2a}
\eeqa
\end{lemma}

\begin{lemma}\label{lm:jin}
With $A>B>0$, let us define $\calC\define [A-B,A+B]$. Then, for all $x,y,z\in\bbR$, the following two statements are equivalent to each other:
\beqa
z &\in&  A \sgn(x-z) + B\sgn(y-z) \dfeq{stt1}\\
z &=& \proj([\proj(-\calC,y),\proj(\calC,y)],x). \dfeq{stt2}
\eeqa
\end{lemma}

\begin{rem}
Lemma~\ref{lm:jin} has been presented by Jin et al.\cite{Jin_2014_AR} but a strict proof had not been presented.
\end{rem}

\section{Continuous Twisting Algorithms and Its Implementation}
\label{sec:control}
To solve the double integrator with significant uncertainty is given as
\begin{align}\label{a1}
\begin{split}
\dot{x}_1&=  x_2\\
\dot{x}_2 & \in  u+\delta(t)
\end{split}
\end{align}
where $\delta(t)\in \mathbb{R}$ represents the matched exogenous/model uncertainty. A design of continuous  twisting algorithm as proposed in \cite{CTA_2015_conference} follows:
\begin{align}\label{a2}
\begin{split}
u & = -\mathcal{L}^{\frac{2}{3}}k_1\lfloor x_1\rceil^{\frac{1}{3}} -\mathcal{L}^{\frac{1}{2}}k_1\lfloor x_2\rceil^{\frac{1}{2}} +\eta\\
\dot{\eta} & \in -\mathcal{L}k_3\lfloor x_1\rceil^{0} -\mathcal{L}k_4\lfloor x_2\rceil^{0}
\end{split}
\end{align}
where $k_i>0,~i=1,2,3,4.$ This control is able to drive the system trajectories of \eqref{a1} to the origin in finite-time and reject the bounded derivative of disturbance $\dot{\delta}(t)= \Delta(t)$ and $|\Delta(t)|\leq L$ with some unknown positive constant. By putting \eqref{a2} in \eqref{a1}, we get
\begin{align}\label{a3}
\begin{split}
\dot{x}_1&= x_2\\
\dot{x}_2 & = -\mathcal{L}^{\frac{2}{3}}k_1\lfloor x_1\rceil^{\frac{1}{3}} -\mathcal{L}^{\frac{1}{2}}k_1\lfloor x_2\rceil^{\frac{1}{2}} +\eta+\delta(t)\\
\dot{\eta} & \in -\mathcal{L}k_3\lfloor  x_1\rceil^{0} -\mathcal{L}k_4\lfloor x_2\rceil^{0}
\end{split}
\end{align}
Although, a fictions state is defined with the second equation of \eqref{a3} as
\begin{align}\label{a4}
x_3 = \eta+\delta(t)
\end{align}
Therefore, the 3-order dynamical system for closed-loop system is given by
\begin{align}  \label{a5}
\begin{split}
\dot{x}_1&= x_2\\
\dot{x}_2 & = -\mathcal{L}^{\frac{2}{3}}k_1\lfloor  x_1\rceil^{\frac{1}{3}} -\mathcal{L}^{\frac{1}{2}}k_1\lfloor  x_2\rceil^{\frac{1}{2}} +x_3\\
\dot{x}_3 & \in  -\mathcal{L}k_3\lfloor x_1\rceil^{0} -\mathcal{L}k_4\lfloor x_2\rceil^{0} +\Delta(t)
\end{split}
\end{align}
The third line of \eqref{a5} may be associated with the differential inclusion due to sign function with disturbance term i.e., $\dot{x}_3 \in -\mathcal{L}k_3\lfloor x_1\rceil^{0} -\mathcal{L}k_4\lfloor  x_2\rceil^{0} +[-L,L].$ The solution is homogeneous in nature of degree $k=-1$ with weights $r=[3,2,1]^{\top}.$ However in nominal case i.e., $\delta(t)=0$, the system trajectory $x_3$ is similar to $\eta$ for all $t\ge 0$.

\par

Let us scale the state $x$:
\begin{eqnarray}
z_{1}=\mathcal{L} x_{1}, \quad z_{2}=\mathcal{L} x_{2}, \quad z_{3}=\mathcal{L} x_{3}
\end{eqnarray}
with a new constant $0<\mathcal{L}\in \bbR$. Then the system \eqref{a5} changes into the following new system:
\begin{align}  \label{state_system_z}
\begin{split}
 \dot{z}_{1} &=z_{2} \\
 \dot{z}_{2} &=-k_{p 1}\left\lceil z_{1}\right\rfloor^{\frac{1}{3}}-k_{p 2}\left\lceil z_{2}\right\rfloor^{\frac{1}{2}}+z_{3} \\
 \dot{z}_{3} &=-k_{p 3}\left\lceil z_{1}\right\rfloor^{0}-k_{p 4}\left[z_{2}\right\rfloor^{0}+{\Delta}_{p}(t)
\end{split}
\end{align}
where
\begin{eqnarray}
\begin{array}{rl}
{k_{p 1}=\mathcal{L}^{\frac{2}{3}} k_{1}} & {k_{p 2}=\mathcal{L}^{\frac{1}{2}} k_{2}} \\ {k_{p 3}=\mathcal{L} k_{3}} & {k_{p 4}=\mathcal{L} k_{4}}.
\end{array}
\end{eqnarray}

\par

To implement the continuous twisting algorithm, some discretization methods can be applied to \eqref{state_system_z}. Let us recall some of them.
Consider the discrete-time formulation of \eqref{a1} in one of traditional Euler's approaches as follows:
\begin{align}\label{a8}
\begin{split}
z_{1,k+1}&= z_{1,k}+hz_{2,k}\\
z_{2,k+1} & = z_{2,k}-hk_{p1}\lfloor  x_{1,k}\rceil^{\frac{1}{3}} -hk_{p2}\lfloor  x_{2,k}\rceil^{\frac{1}{2}} +hz_{3,k}\\
z_{3,k+1} & = z_{3,k}- hk_{p3}\lfloor  x_{1,k}\rceil^{0} -hk_{p4}\lfloor  z_{2,k}\rceil^{0}+h\Delta_{p,k}
\end{split}
\end{align}
where $x_{i,k}:=x_i(t_k)$, $h=t_{k+1}-t_k>0$ represents time step interval and the disturbance term $\Delta_{p,k}$ is expressed as
\begin{align}\label{a11}
\delta_{k+1}=\delta_k+h\Delta_{p,k}.
\end{align}
The above discrete-time formulation \eqref{a8} is an approximation of \eqref{a1}\eqref{a2}, which can leads to different properties from the continuous one. Here, alternative ways for relatively smoother discrete-time results are conducted with the implementation of implicit-Euler approach.

\section{Proposed Implicit Euler Implementation}
Before introducing the proposed implicit Euler implementation of \eqref{a8}, let us rewrite it as:
\begin{align}\label{z_model}
\begin{split}
\dot{z}_1&=  z_2\\
\dot{z}_2 & =  u_1+\eta+\delta(t)
\end{split}
\end{align}
where $u_1=-k_{p 1}\left\lceil z_{1}\right\rfloor^{\frac{1}{3}}-k_{p 2}\left\lceil z_{2}\right\rfloor^{\frac{1}{2}}$ and $\eta$ is an intermediate variable used to compensate the disturbance $\delta(t)$. It is assumed that both $\eta$ and $\delta(t)$ are differentiable. Their derivatives are $\dot \eta=-k_{p 3}\left\lceil z_{1}\right\rfloor^{0}-k_{p 4}\left[z_{2}\right\rfloor^{0}$ and $\dot \eta=\Delta(t)$, respectively.

\par

The implicit Euler discretization of \eqref{z_model} is as follows:
\begin{align}\label{z_model_discretized}
\begin{split}
{z}_{1,k+1}&=  z_{1,k}+h{z}_{2,k+1}\\
{z}_{2,k+1} & = z_{2,k}+h u_{1,k}+h\eta_{k+1}+h\delta_{k+1}
\end{split}
\end{align}
where
\begin{align}\label{u_discrete}
\begin{split}
u_{1,k}& \in -k_{p 1}\left\lceil z_{1,k+1}\right\rfloor^{\frac{1}{3}}-k_{p2}\left\lceil z_{2,k+1}\right\rfloor^{\frac{1}{2}}.
\end{split}
\end{align}
The proposed implicit Euler implementation is divided into two stages and both stages are based on the discrete system \eqref{z_model_discretized}.

\subsection{Stage I}

 The first stage begins with the normal version of the discrete system \eqref{z_model_discretized}:
\begin{align}\label{z_model_normal}
\begin{split}
\tilde{z}_{1,k+1}&=  z_{1,k}+h\tilde{z}_{2,k+1}\\
\tilde{z}_{2,k+1} & = z_{2,k}+h u_{1,k}
\end{split}
\end{align}
where $u_{1,k}$ is the normal version:
\begin{align}\label{u_normal_discrete}
\begin{split}
u_{1,k}& \in -k_{p, 1}|\bar z_{1,k}|^{\frac{1}{3}}\mathrm{sgn}(\tilde{z}_{1,k+1})-k_{p,2}  |\bar {z}_{2,k}|^{\frac{1}{2}}\mathrm{sgn}(\tilde{z}_{2,k+1})
\end{split}
\end{align}
with intermediate variables $\tilde{z}_{i,k+1}$ and $\bar z_{i,k}$, $i\in \{1,2\}$. Submitting \eqref{u_normal_discrete} into \eqref{z_model_normal}, one has the following expression:
\begin{align}\label{z_normal_closed_loop_01}
\begin{split}
\tilde{z}_{1,k+1}&=  z_{1,k}+h\tilde{z}_{2,k+1}\\
\tilde{z}_{2,k+1} & \in z_{2,k}  -k_{p, 1}|\bar z_{1,k}|^{\frac{1}{3}}\mathrm{sgn}(\tilde{z}_{1,k+1}) \\
\qquad &-k_{p,2}  |\bar {z}_{2,k}|^{\frac{1}{2}}\mathrm{sgn}(\tilde{z}_{2,k+1}),
\end{split}
\end{align}
from which one has:
\begin{align}\label{z_normal_closed_loop_02}
\begin{split}
\tilde{z}_{2,k+1} & \in z_{2,k}  -k_{p, 1}|\bar z_{1,k}|^{\frac{1}{3}}\mathrm{sgn}\left(\frac{z_{1,k}}{h}+\tilde{z}_{2,k+1}\right)\\
\qquad &  -k_{p,2}  |\bar {z}_{2,k}|^{\frac{1}{2}}\mathrm{sgn}(\tilde{z}_{2,k+1}).
\end{split}
\end{align}
By applying Lemma~\ref{lm:jin}, one has:
\begin{eqnarray}
\tilde{z}_{2,k+1}&= &\mathrm{proj} \bigg([\mathrm{proj}(-\mathcal{A}_{k},-z_{2,k}),\nonumber \\
&&\qquad  \mathrm{proj}(\mathcal{A}_{k},-z_{2,k})],  -z_{2,k}-\frac{z_{1,k}}{h}\bigg)
\end{eqnarray}
where $\mathcal{A}_{k}\define [(k_{p1}|\bar{z}_{1,k}|^\frac{1}{3}-k_{p2}|\bar{z}_{2,k}|^\frac{1}{2}),(k_{p1}|\bar{z}_{1,k}|^\frac{1}{3}+k_{p2}|\bar{z}_{2,k}|^\frac{1}{2})]$.
In conclusion, one has:
\begin{subequations}\label{1_level_conclusion}
\begin{eqnarray}
u_{1,k}&=&\frac{1}{h}\mathrm{proj} \bigg([\mathrm{proj}(-\mathcal{A}_{k},-z_{2,k}),\nonumber \\
&&\qquad  \mathrm{proj}(\mathcal{A}_{k},-z_{2,k})],  -z_{2,k}-\frac{z_{1,k}}{h}\bigg)\\
\tilde{z}_{2,k+1}&= & z_{2,k}+hu_{1,k} \\
\tilde{z}_{1,k+1}&=&z_{1,k}+h\tilde{z}_{2,k+1}.
\end{eqnarray}
\end{subequations}

\subsection{Stage II}

Let us introduce the second stage of the proposed implementation. One can rewrite \eqref{z_model_discretized} as follows:
\bsbeq  \label{state_discrete_02}
\begin{align}
z_{1,k+1}&= z_{1,k}+hz_{2,k+1}\\
z_{2,k+1} &= z_{2,k}+hu_{1,k}+h\eta_{k+1}+h\delta_{k+1}.
\end{align}
\esbeq
where $u_{1,k}$ is obtained from \eqref{1_level_conclusion}. Another control input $\eta_{k+1}$ is discretized by using the implicit Euler method as follows:
\beqa \label{u_disc}
\eta_{k+1} & \in & \eta_k-hk_{p,3}\lfloor  \bar{z}_{1,k+1}\rceil^{0} -hk_{p,4}\lfloor  \bar{z}_{2,k+1}\rceil^{0}
\eeqa
 where $\bar{z}_{i,k+1}$, $i\in\{1,2\}$ are the intermediate variables and its updating will be introduced later.

 \par

Let us introduce $z_{3,k+1}\define \eta_{k+1}+\delta_{k+1}$. Then \eqref{state_discrete_02} can be rewrite as:
\bsbeq  \label{rewrite_z}
\begin{align}
z_{1,k+1}&= z_{1,k}+hz_{2,k+1}\\
z_{2,k+1} &= z_{2,k}+hu_{1,k}+h z_{3,k+1}\\
z_{3,k+1} &= z_{3,k}-hk_{p3}\lfloor  \bar{z}_{1,k+1}\rceil^{0} -hk_{p4}\lfloor  \bar{z}_{2,k+1}\rceil^{0}+h\Delta_k
\end{align}
\esbeq
where $z_{3,k}\define \eta_k+\delta_k$ and $\delta_{k+1}=\delta_k+h\Delta_k$.

Here, in a similar way, we first consider the corresponding normal version of \eqref{rewrite_z} as follows:
\bsbeq  \label{normal_mixed}
\begin{align}
{\bar {z}_{1,k+1}}&= z_{1,k}+h \bar{z}_{2,k+1}\\
\bar{z}_{2,k+1}&=z_{2,k}+hu_{1,k}+h\bar{z}_{3,k+1}\\
{\bar {z}_{3,k+1}}&=z_{3,k}-hk_{p3}\lfloor  \bar{z}_{1,k+1}\rceil^{0} -hk_{p4}\lfloor  \bar{z}_{2,k+1}\rceil^{0}
\end{align}
\esbeq

%

\begin{figure*}
\centering
\begin{minipage}[b]{.48\textwidth}
\includegraphics[width=1.1\textwidth]{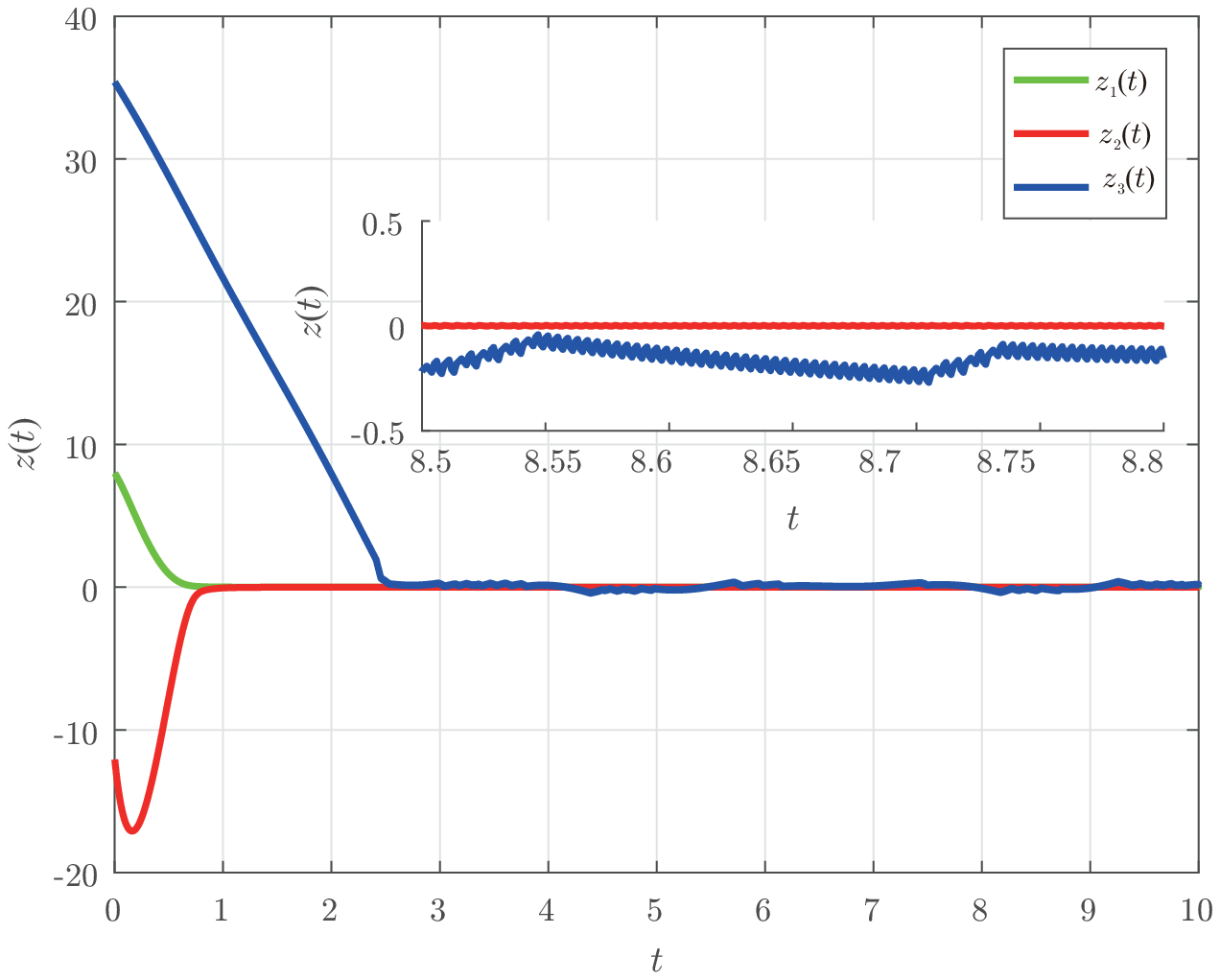}
\caption{Convergence of states $z_1, z_2$ and $z_3$ with the conventional explicit Euler method \eqref{explicit_euler}.}\label{fig_1}
\end{minipage}\qquad
\begin{minipage}[b]{.48\textwidth}
\includegraphics[width=1.1\textwidth]{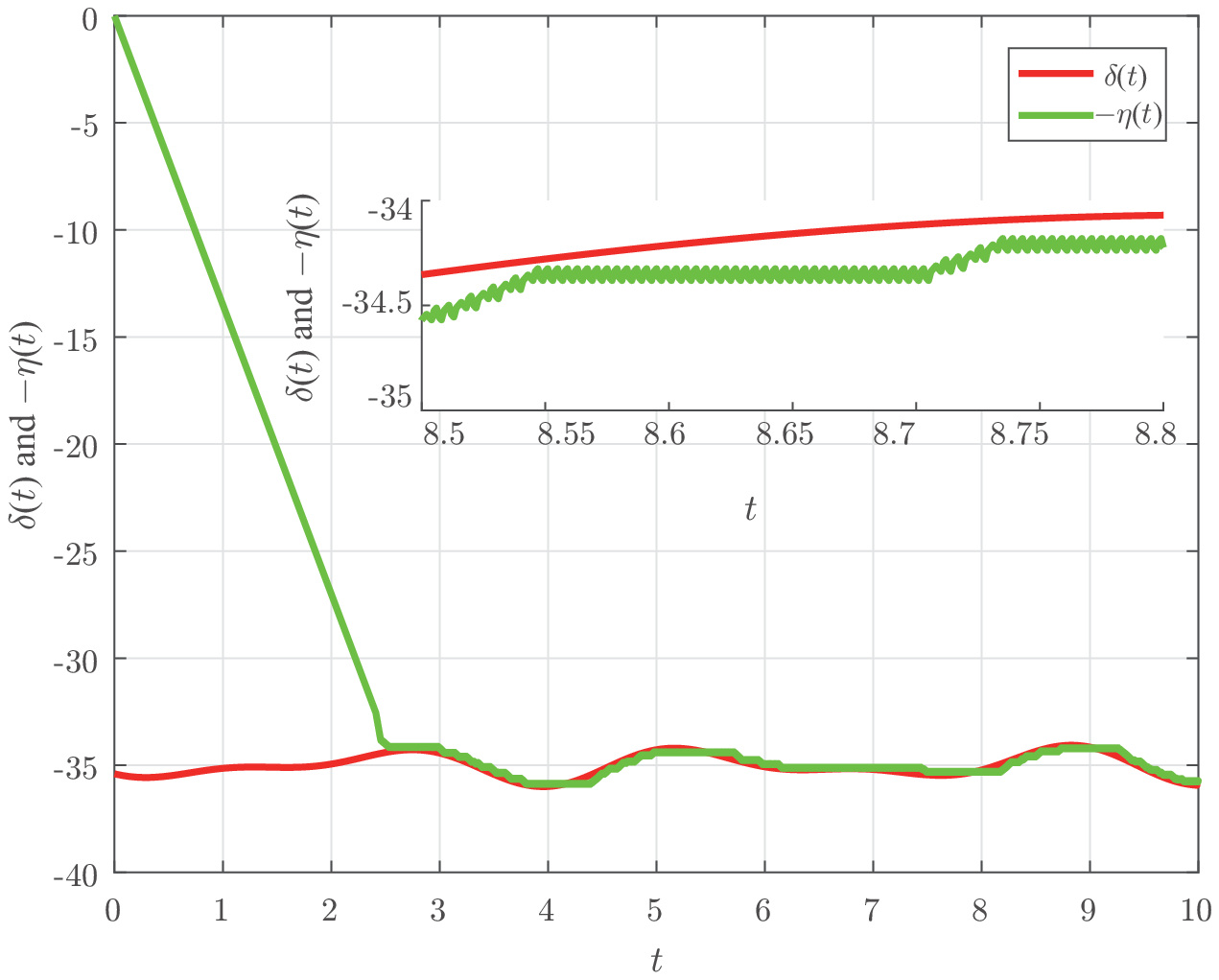}
\caption{Convergence of $-\eta$ to the disturbance $\delta(t)$ with the conventional explicit Euler method \eqref{explicit_euler}}\label{fig_2}
\end{minipage}
\end{figure*}

\begin{figure*}
\centering
\begin{minipage}[b]{.48\textwidth}
\includegraphics[width=1.1\textwidth]{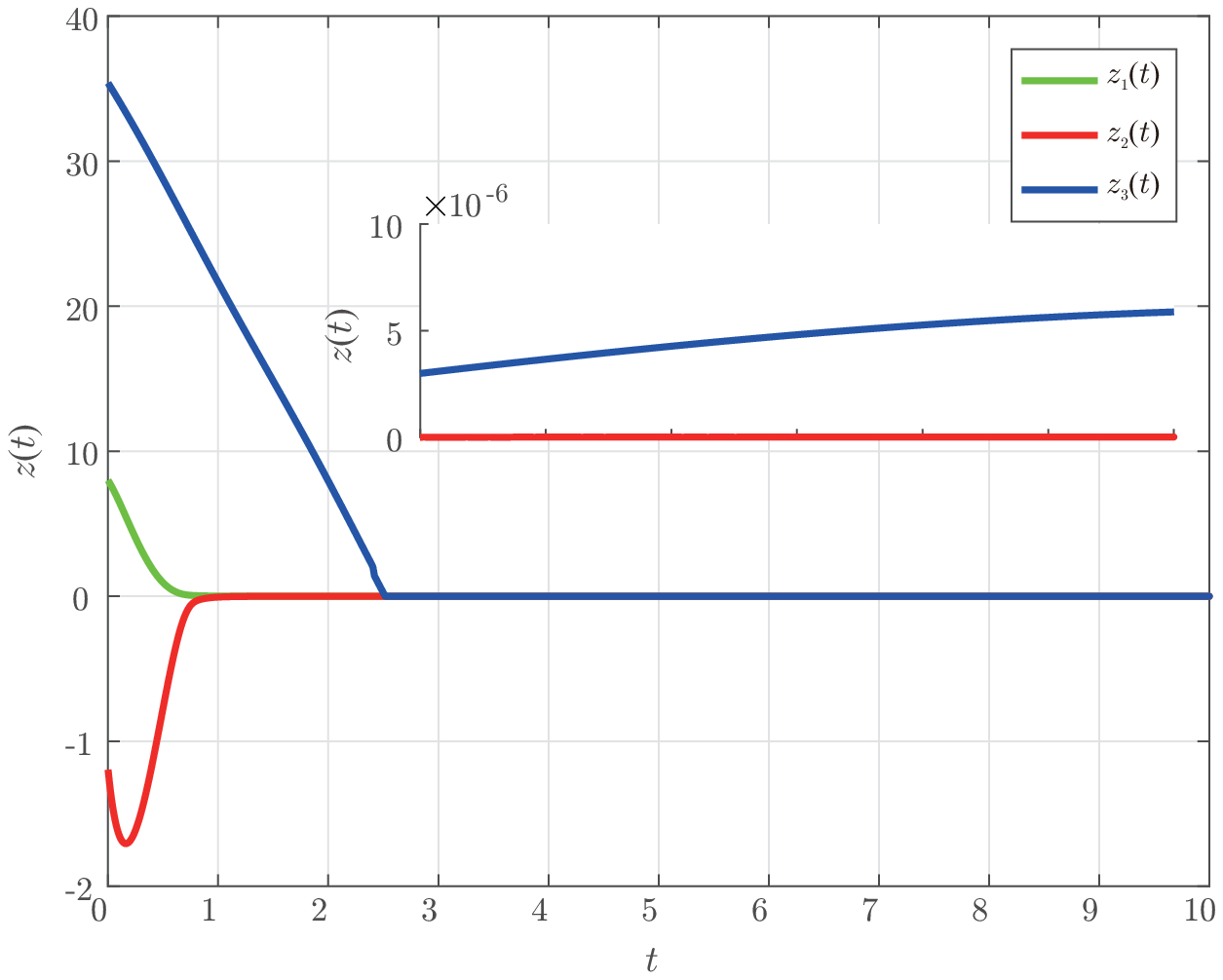}
\caption{Convergence of states $z_1, z_2$ and $z_3$ with the proposed implicit Euler method \eqref{u_k_form}.}\label{fig_3}
\end{minipage}\qquad
\begin{minipage}[b]{.48\textwidth}
\includegraphics[width=1.1\textwidth]{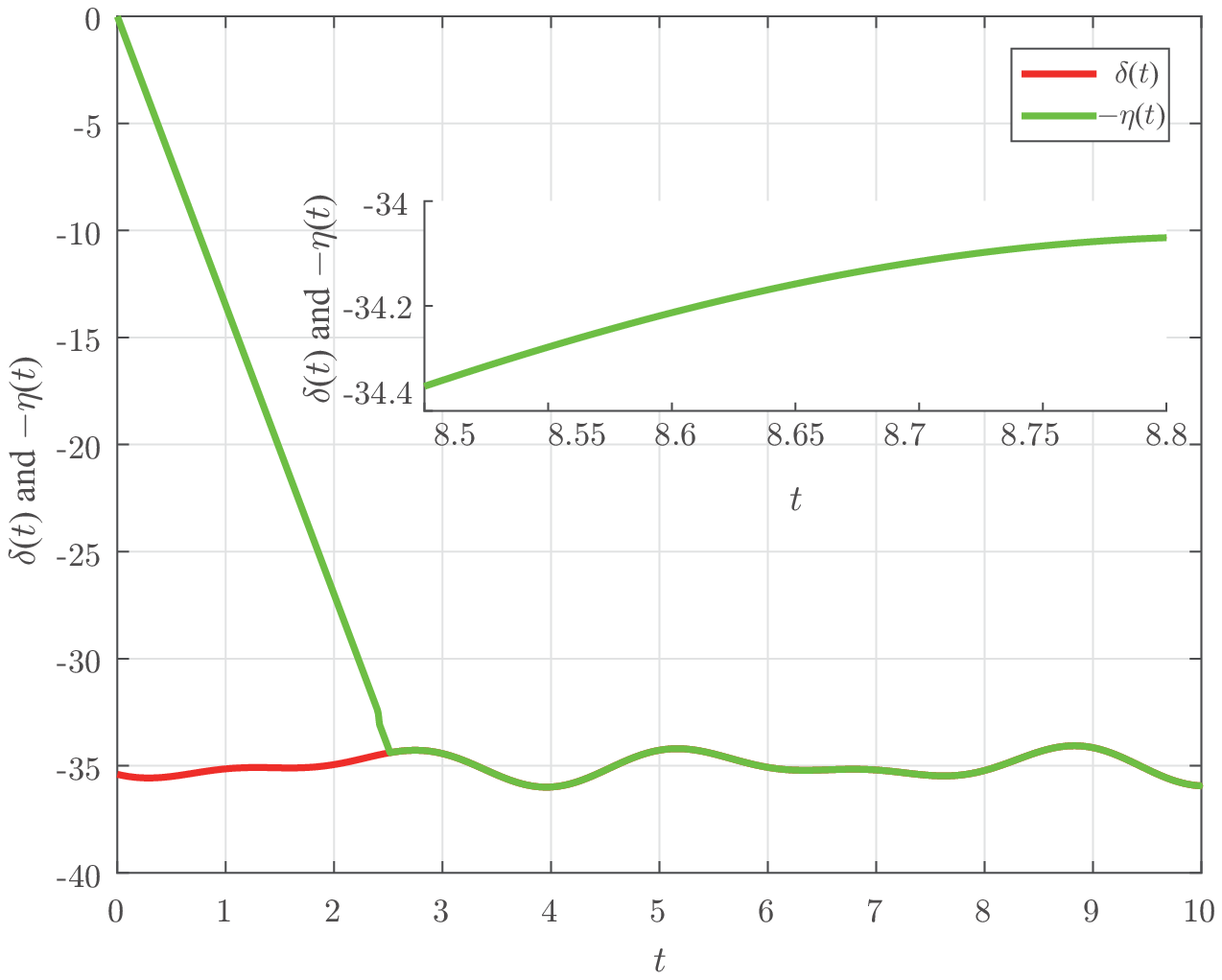}
\caption{Convergence of $-\eta$ to the disturbance $\delta(t)$ with the proposed implicit Euler method \eqref{u_k_form}}\label{fig_4}
\end{minipage}
\end{figure*}

From \eqref{normal_mixed}(a)-(b), one has:
\bsbeq  \label{linear_relation}
\begin{align}
\bar{z}_{2,k+1}&= \left(\frac{z_{2,k}+hu_{1,k}}{h} +\bar{z}_{3,k+1}\right)h \\
\bar{z}_{1,k+1}&=  \left(\frac{z_{1,k}+hz_{2,k}+h^2u_{1,k}}{h^2} +\bar{z}_{3,k+1}\right)h^2.
\end{align}
\esbeq
Substituting $\bar{x}_{2,k+1}$ and $\bar{x}_{1,k+1}$ from \eqref{linear_relation} into \eqref{normal_mixed}(c), we can derive the reduced form of the following inclusion as:
\bsbeq
\begin{align}    \label{mix_x_2}
\bar{z}_{3,k+1} & \in z_{3,k} -hk_{p,3} \lfloor  \frac{z_{1,k}}{h^2}+\frac{z_{2,k}+hu_{1,k}}{h} +\bar{z}_{3,k+1} \rceil^0 \\
&-hk_{p,4} \lfloor \frac{z_{2,k}+hu_{1,k}}{h} +\bar{z}_{3,k+1}\rceil^0.
\end{align}
\esbeq
By following the Lemma~\ref{lm:jin}, the solution of \eqref{mix_x_2} can be obtained with as follows:
\begin{align} \label{solution_implicit}
\begin{split}
\bar{z}_{3,k+1}&= z_{3,k} + \mathrm{proj} \bigg([\mathrm{proj}(-\mathcal{B}_{k},-y_{1,k}),\\
&\mathrm{proj}(\mathcal{B}_{k},-y_{1,k})],  -y_{2,k}\bigg)
\end{split}
\end{align}
where $\mathcal{B}_{k}\define [h(k_{p,3}-k_{p,4}), h(k_{p,3}+h_{p,4})]$ and
\bsbeq
\begin{align} \label{y1_y2}
y_{1,k}&\define  \frac{z_{2,k}+hu_{1,k}}{h}+z_{3,k} \\
y_{2,k}&\define  \frac{z_{1,k}}{h^2}+\frac{z_{2,k}+hu_{1,k}}{h}+z_{3,k}.
\end{align}
\esbeq
Then, the intermediate variables $\bar{z}_{i,k+1},i\in\{1,2,3\}$ are updated as follows:
\begin{subequations}\label{z_i_update}
\begin{eqnarray}
y_{1,k}&:=& \frac{z_{2,k}+hu_{1,k}}{h}+z_{3,k} \\
y_{2,k}&:=& \frac{z_{1,k}}{h^2}+\frac{z_{2,k}+hu_{1,k}}{h}+z_{3,k} \\
\bar{z}_{3,k+1}&=& z_{3,k} + \mathrm{proj} \bigg([\mathrm{proj}(-\mathcal{B}_{k},-y_{1,k}),\nonumber \\
&&\qquad \mathrm{proj}(\mathcal{B}_{k},-y_{1,k})],  -y_{2,k}\bigg) \\
\bar{z}_{2,k+1}&=&\left(\frac{z_{2,k}+hu_{1,k}}{h} +\bar{z}_{3,k+1}\right)h \\
\bar{z}_{1,k+1}&=&\left(\frac{z_{1,k}+h^2z_{2,k}+hu_{1,k}}{h^2} +\bar{z}_{3,k+1}\right)h^2.
\end{eqnarray}
\end{subequations}
Finally, the implementation of the CTA controller $u_k:=u_{1,k}+\eta_{k+1}$ is as follows:
\begin{subequations}\label{u_k_form}
\begin{eqnarray}
u_{1,k}&=&\frac{1}{h}\mathrm{proj} \bigg([\mathrm{proj}(-\mathcal{A}_{k},-z_{2,k}),\nonumber \\
&&\qquad  \mathrm{proj}(\mathcal{A}_{k},-z_{2,k})],  -z_{2,k}-\frac{z_{1,k}}{h}\bigg)\\
\eta_{k+1}&:=& \eta_k+ \mathrm{proj} \bigg([\mathrm{proj}(-\mathcal{B}_{k},-y_{1,k}),\nonumber \\
&&\qquad \mathrm{proj}(\mathcal{B}_{k},-y_{1,k})],  -y_{2,k}\bigg) \\
u_k&=&u_{1,k}+\eta_{k+1}
\end{eqnarray}
\end{subequations}

%
%
%

\begin{figure*}
\centering
\begin{minipage}[b]{.48\textwidth}
\includegraphics[width=1.1\textwidth]{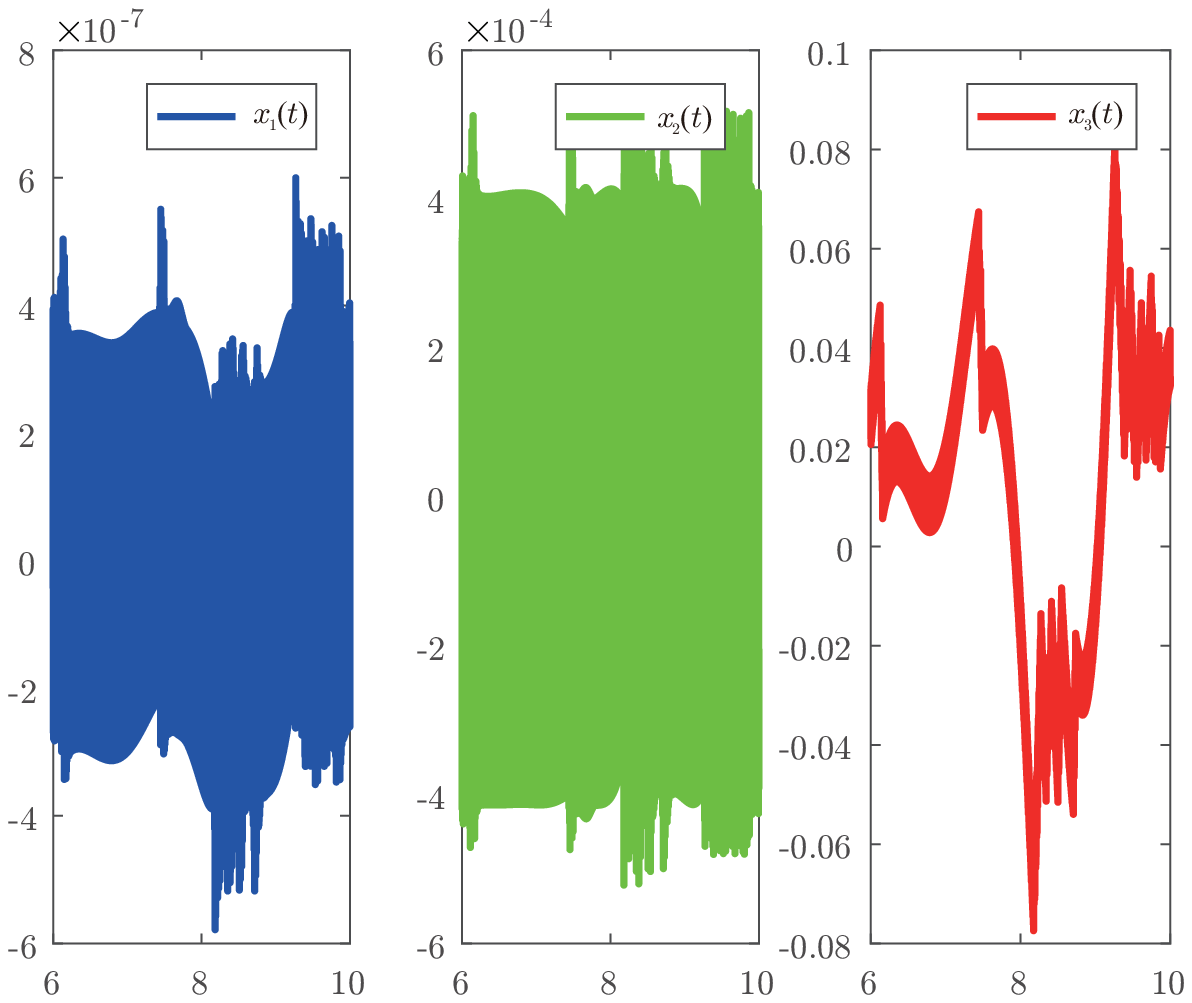}
\caption{Precision of the state variables $x_1, x_2$ and $x_3$ with the conventional explicit Euler method \eqref{explicit_euler} and the time step size is $h=0.001$s.}\label{fig_5}
\end{minipage}\qquad
\begin{minipage}[b]{.48\textwidth}
\includegraphics[width=1.1\textwidth]{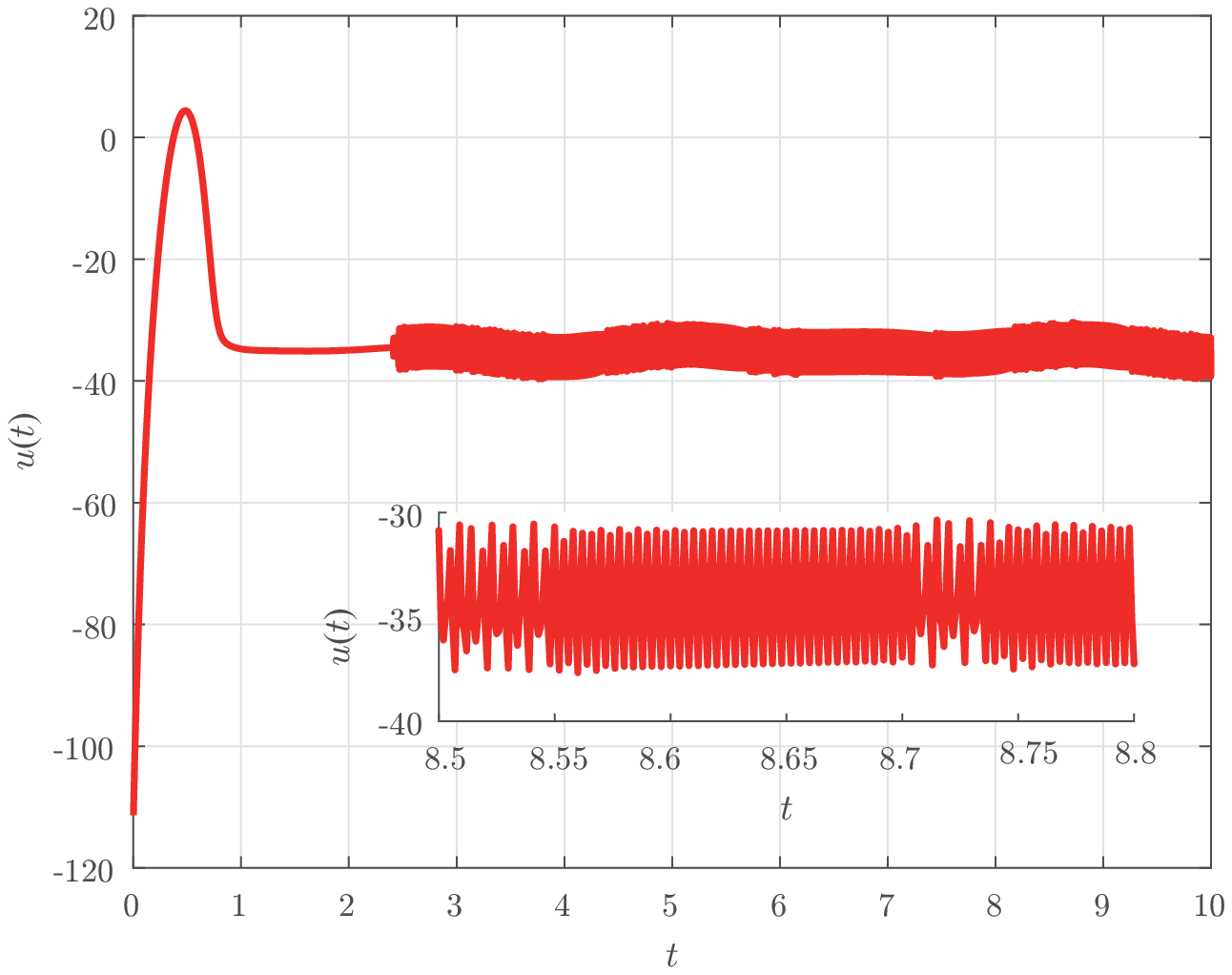}
\caption{The control input signal $u(t)$ applied to \eqref{simu_01} with the conventional explicit Euler method \eqref{explicit_euler} and the disturbance is set as $\eta(t)=35+0.6\mathrm{sin}(2t)+0.4\mathrm{sin}(\sqrt{10}t)$.}\label{fig_6}
\end{minipage}
\end{figure*}

\begin{figure*}
\centering
\begin{minipage}[b]{.48\textwidth}
\includegraphics[width=1.1\textwidth]{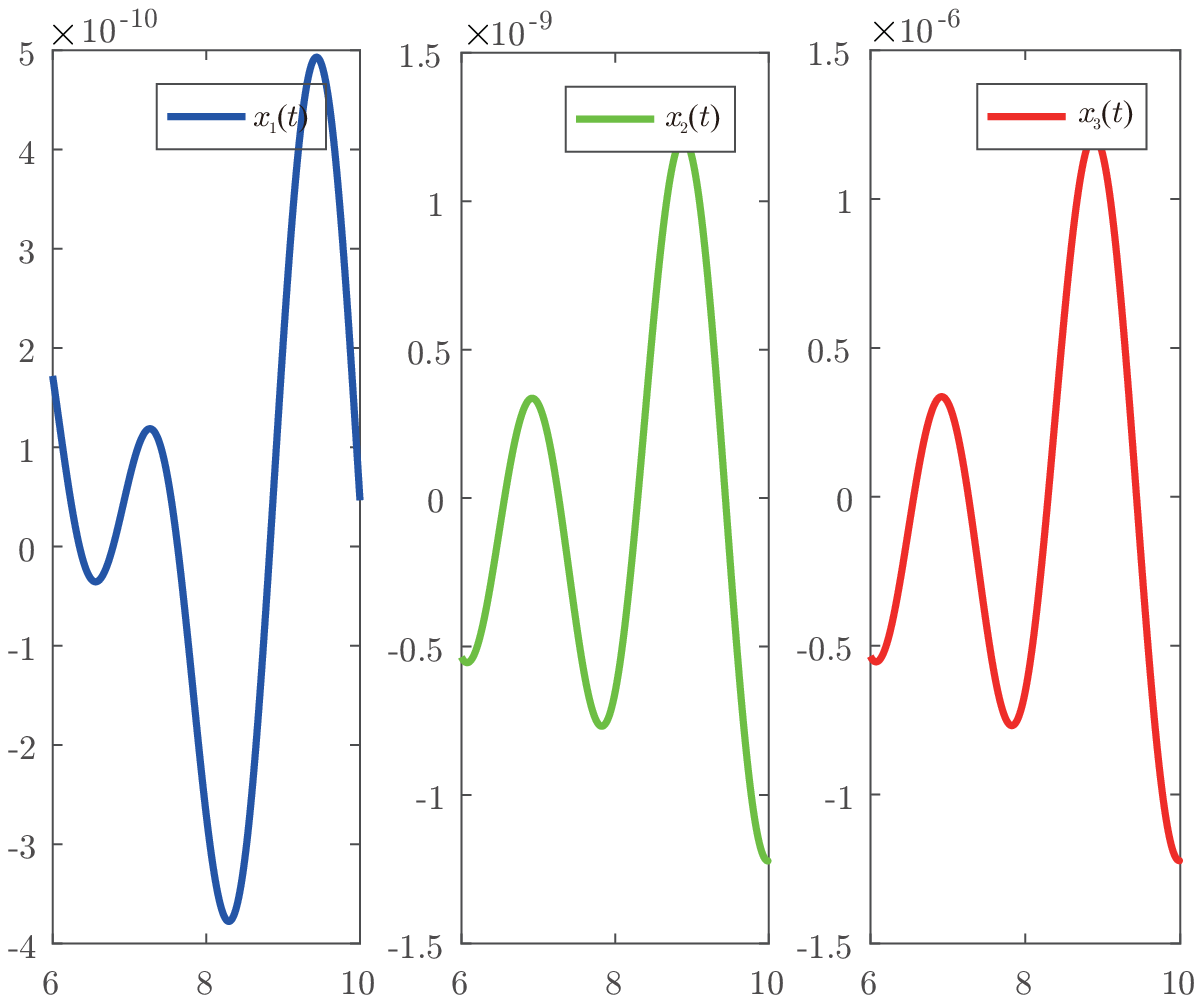}
\caption{Precision of the state variables $x_1, x_2$ and $x_3$ with the proposed implicit Euler method \eqref{u_k_form} and the time step size is $h=0.001$s.}\label{fig_7}
\end{minipage}\qquad
\begin{minipage}[b]{.48\textwidth}
\includegraphics[width=1.1\textwidth]{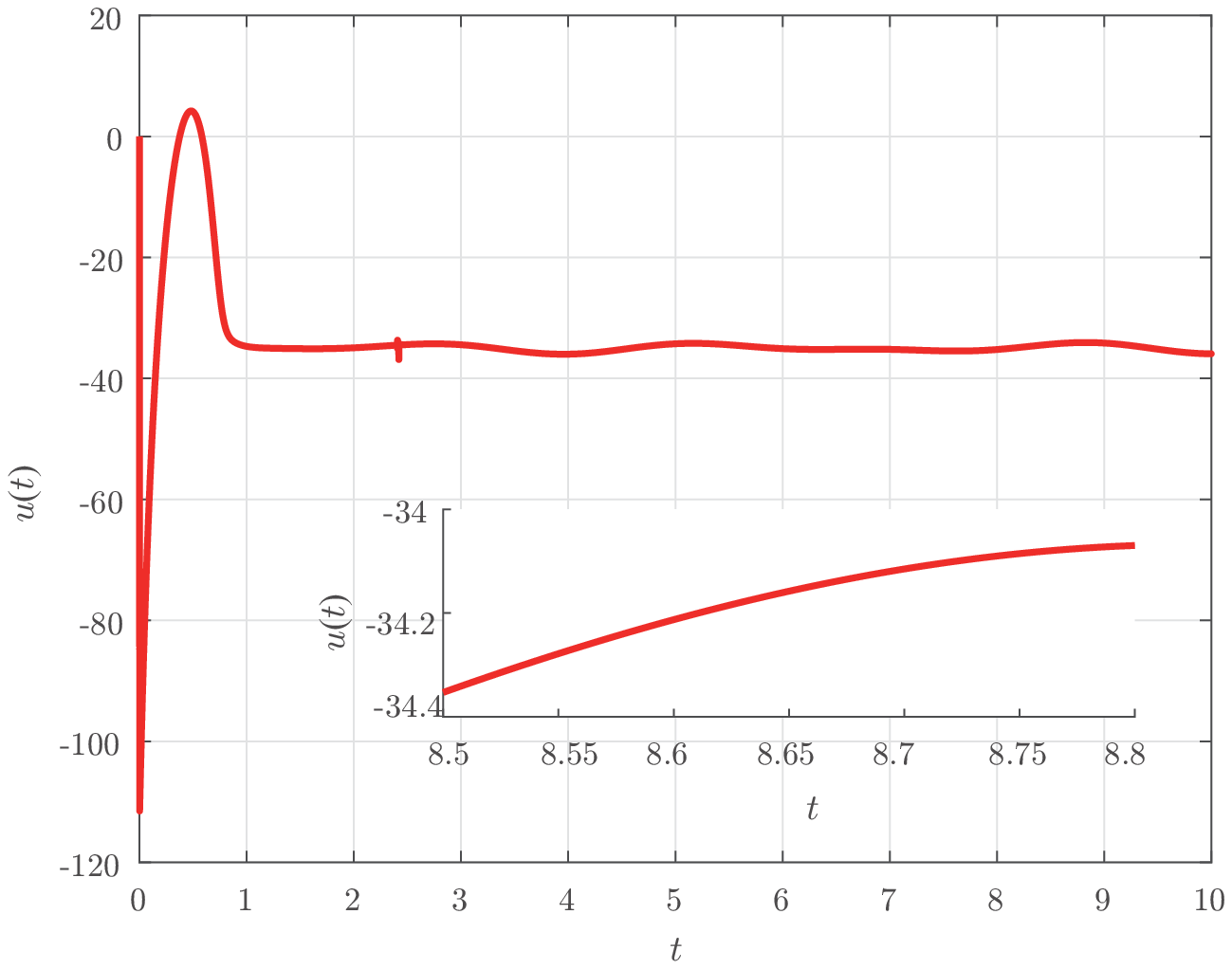}
\caption{The control input signal $u(t)$ applied to \eqref{simu_01} with the proposed implicit Euler method \eqref{u_k_form} and the disturbance is set as $\eta(t)=35+0.6\mathrm{sin}(2t)+0.4\mathrm{sin}(\sqrt{10}t)$.}\label{fig_8}
\end{minipage}
\end{figure*}

\section{SIMULATIONS}\label{sec:simulation}
The proposed algorithms are compared with the conventional explicit Euler implementations through some simulations. The time evolution of the states of system \eqref{a1} is illustrated in Fig.1. The disturbance is set as $\delta(t)=35 + 0.6\mathrm{cos}(2t) + 0.4 \mathrm{sin}(\sqrt{10}t)$ and its derivation is $\Delta(t)=1.2\mathrm{cos}(2t) + 0.4 \sqrt{10}\mathrm{sin}(\sqrt{10}t)$. The gains for the CTA that achieve finite-time convergence to zero of $z_1$ and $z_2$ are
\begin{align*}
 k_{p,1}&=160.236, \quad k_{p,2}=60.3738 \\
k_{p,3}&=28.5, \quad k_{p,4}=15
\end{align*}
and the scaling factor $L=5$.

\par

In simulations of the conventional implementation and the proposed implementation of the CTA, the following discrete system is employed:
 \begin{align}\label{simu_01}
\begin{split}
z_{1,k+1}&= z_{1,k}+hz_{2,k}\\
z_{2,k+1} & = z_{2,k}+hu_k+h\delta_k\\
z_{3,k+1} & = \eta_{k}+\delta_k.
\end{split}
\end{align}
where $\delta_k=35 + 0.6\mathrm{cos}(2t(k)) + 0.4 \mathrm{sin}(\sqrt{10}t(k))$. In the simulation of the conventional explicit Euler method, the following expression is used:
\begin{subequations}\label{explicit_euler}
\begin{eqnarray}
u_k&=&u_{1,k}+\eta_k\\
u_{1,k}&=&-hk_{p1}\lfloor  x_{1,k}\rceil^{\frac{1}{3}} -hk_{p2}\lfloor  x_{2,k}\rceil^{\frac{1}{2}} +h\eta_k \\
\eta_{k+1}&=&\eta_k- hk_{p3}\lfloor  x_{1,k}\rceil^{0} -hk_{p4}\lfloor  z_{2,k}\rceil^{0},
\end{eqnarray}
\end{subequations}
while in the simulation of the proposed implicit Euler method, the expression \eqref{u_k_form} is employed. In both cases, the time step size is chosen as $h=0.001$s and the initial states are set as: $z_1(0)=8$, $z_2(0)=-12$ and $z_3(0)=35$.


 The states $z_1$, $z_2$, and $z_3$ with the conventional explicit Euler method are shown in Fig.~\ref{fig_1}. Fig.~\ref{fig_1} shows that the states $z_1$, $z_2$ and $z_3$ converges to zeros in finite time. However, the inset plot in Fig.~\ref{fig_1} shows that $z_3$ deviates from zero a little and the chattering appears with a small magnitude in $z_3$. Similar phenomenons happen in the inset plot of Fig.~\ref{fig_2}. In Fig.~\ref{fig_2}, the variable $\eta$ compensates the disturbance $\delta(t)$, i.e., $-\eta=\delta(t)$. However, the inset plot in Fig.~\ref{fig_2} shows that $-\eta$ deviates from $\delta(t)$ a little and a small of magnitude of chattering happens in $\eta$.

 \par

  The states $z_1$, $z_2$, and $z_3$ with the proposed implicit Euler method are shown in Fig.~\ref{fig_3}. Fig.~\ref{fig_3} shows that the states $z_1$, $z_2$ and $z_3$ converges to zeros in finite time. The time of convergence is the same as that of the conventional explicit Euler method, about $2.5$s. This means that the proposed implicit discretization does not change the convergence property of the CTA. Furthermore, the inset plot of Fig.~\ref{fig_3} shows that $z_3$ converges to zero with a precision about $10^{-6}$ and there is no chattering. One can also observe that $\eta(t)$ compensates $\delta(t)$ with a high precision and no chattering happens.

\par

The precision of the system states before scaling, i.e., $x_i=z_i/L, i\in\{1,2,3\}$, is shown Fig.~\ref{fig_5}. Within the period $8\leq t\leq 10$s, Fig.~\ref{fig_5} shows that the precision of $x_i$ is about
 \begin{align}\label{simu_03}
\begin{split}
|x_1|\leq v_1 h^3, |x_2|\leq v_2 h^2, |x_3|\leq v_3 h
\end{split}
\end{align}
with constants
 \begin{align}\label{simu_04}
\begin{split}
v_1=600 , v_2=610, v_3=80.
\end{split}
\end{align}

One can also note that the proposed implicit Euler method achieve a much higher precision, as shown in Fig.~\ref{fig_5}. It satisfies the following inequalities:
 \begin{align}\label{simu_05}
\begin{split}
|x_1|\leq v_1 h^4, |x_2|\leq v_2 h^3, |x_3|\leq v_3 h^2
\end{split}
\end{align}
with constants
 \begin{align}\label{simu_06}
\begin{split}
v_1=500 , v_2=1.5, v_3=1.5.
\end{split}
\end{align}

Fig~\ref{fig_6} and Fig.~\ref{fig_8} show the control input $u(t)$ with the conventional explicit Euler and proposed implicit Euler method, respectively. They illustrate that both the methods have almost the same magnitude of control input and this is very important for some actuators. Furthermore, the inset plot in Fig~\ref{fig_6} clearly shows that there is chattering in the control input when the CTA implemented with the conventional explicit Euler method. In contrast, the control input signal calculated by using the proposed implicit Euler method is very smooth, as shown in the inset plot in Fig~\ref{fig_6}. Only one small chattering happens when $z_3$ goes to zero. The smooth input is desired for practical applications.


\section{CONCLUSIONS}\label{sec:conclusion}

This paper proposed implicit-Euler implementations of continuous terminal algorithm (CTA) for second-order systems. Stability properties of the discretized CTA are provided. Trajectory of a typical second-order system was taken as an illustrated example. Comparison shows that the proposed implicit-Euler implementation is superior than the conventional explicit-Euler implementation in term of the suppressions of chattering.

\par

Proofs of finite-time convergence of the discretized by the proposed implicit Euler method may need to be sought in future study. It should also demonstrate the efficiency of the proposed implicit Euler method in real platforms with experiments.

\addtolength{\textheight}{-12cm}   




\section*{ACKNOWLEDGMENT}
This work was finally supported by the National Natural Science Foundation of China (Grant no. 11702073 and Grant No. 61605234), Shenzhen Science and Technology Project (JCYJ20160531174039457), and Shenzhen Key Lab Fund of Mechanisms and Control in Aerospace (Grant no. ZDSYS201703031002066)

\bibliographystyle{IEEEtran}
\bibliography{Mybibfile}

\begin{thebibliography}{10}
\providecommand{\url}[1]{#1}
\csname url@rmstyle\endcsname
\providecommand{\newblock}{\relax}
\providecommand{\bibinfo}[2]{#2}
\providecommand\BIBentrySTDinterwordspacing{\spaceskip=0pt\relax}
\providecommand\BIBentryALTinterwordstretchfactor{4}
\providecommand\BIBentryALTinterwordspacing{\spaceskip=\fontdimen2\font plus
\BIBentryALTinterwordstretchfactor\fontdimen3\font minus
  \fontdimen4\font\relax}
\providecommand\BIBforeignlanguage[2]{{%
\expandafter\ifx\csname l@#1\endcsname\relax
\typeout{** WARNING: IEEEtran.bst: No hyphenation pattern has been}%
\typeout{** loaded for the language `#1'. Using the pattern for}%
\typeout{** the default language instead.}%
\else
\language=\csname l@#1\endcsname
\fi
#2}}

\bibitem{Davila_2005_Oberver}
J.~Davila, L.~Fridman, and A.~Levant, ``Second-order sliding-mode observer for
  mechanical systems,'' \emph{{IEEE} Trans. on Automatic Control}, vol.~50,
  no.~11, pp. 1785--1789, 2005.

\bibitem{Levant_1993_Order}
A.~Levant, ``Sliding order and sliding accuracy in sliding mode control,''
  \emph{International Journal of Control}, vol.~58, no.~6, pp. 1247--1263,
  1993.

\bibitem{Levant_2007_principle}
------, ``Principles of 2 sliding mode design,'' \emph{Automatica}, vol.~43,
  no.~4, pp. 576--586, 2007.

\bibitem{Shyam_2015_Continuous}
A.~Chalanga, S.~Kamal, and B.~Bandyopadhyay, ``A new algorithm for continuous
  sliding mode control with implementation to industrial emulator setup,''
  \emph{IEEE/ASME Transaction on mechatronics}, vol.~20, no.~5, pp. 2194--2204,
  2015.

\bibitem{Xiong_2018_Adaptive01}
D.~Luo, X.~Xiong, S.~Kamal, and S.~Jin, ``\BIBforeignlanguage{English}{Adaptive
  gains of dual level to super-twisting algorithm for sliding mode design},''
  \emph{\BIBforeignlanguage{English}{IET Control Theory \& Applications}},
  vol.~12, pp. 2347--2356(9), November 2018.

\bibitem{Xiong_2018_Adaptive02}
\BIBentryALTinterwordspacing
X.~Xiong, S.~Kamal, and S.~Jin, ``Adaptive gains to super-twisting technique
  for sliding mode design,'' \emph{CoRR}, vol. abs/1805.07761, 2018. [Online].
  Available: \url{http://arxiv.org/abs/1805.07761}
\BIBentrySTDinterwordspacing

\bibitem{Taleb_2014_HOSMC}
M.~Taleb, F.~Plestan, and B.~Bououlid, ``An adaptive solution for robust
  control based on integral high-order sliding mode concept,''
  \emph{International Journal of Robust and Nonlinear Control}, vol.~25, no.~8,
  pp. 1201--1213, 2015.

\bibitem{Utkin_2013_ASTA}
V.~I. Utkin and A.~S. Poznyak, ``Adaptive sliding mode control with application
  to super-twist algorithm: {E}quivalent control method,'' \emph{Automatica},
  vol.~49, no.~1, pp. 39--47, 2013.

\bibitem{Acary_2012_Euler}
V.~Acary, B.~Brogliato, and Y.~V. Orlov, ``Chattering-free digital sliding-mode
  control with state observer and disturbance rejection,'' \emph{IEEE Trans. on
  Automatic Control}, vol.~57, no.~5, pp. 1087--1101, 2012.

\bibitem{ACARY2010284}
V.~Acary and B.~Brogliato, ``Implicit {E}uler numerical scheme and
  chattering-free implementation of sliding mode systems,'' \emph{Systems \&
  Control Letters}, vol.~59, no.~5, pp. 284 -- 293, 2010.

\bibitem{Shyam_2016_CTSMC}
S.~Kamal, J.~A. Moreno, A.~Chalanga, B.~Bandyopadhyay, and L.~M. Fridman,
  ``Continuous terminal sliding-mode controller,'' \emph{Automatica}, vol.~69,
  pp. 308 -- 314, 2016.

\bibitem{CTA_2015_conference}
V.~{Torres-Gonz\'{a}lez}, L.~M. {Fridman}, and J.~A. {Moreno}, ``Continuous
  twisting algorithm,'' in \emph{2015 54th IEEE Conference on Decision and
  Control (CDC)}, Dec 2015, pp. 5397--5401.

\bibitem{CTA_2017_journal}
V.~Torres-González, T.~Sanchez, L.~M. Fridman, and J.~A. Moreno, ``Design of
  continuous twisting algorithm,'' \emph{Automatica}, vol.~80, pp. 119 -- 126,
  2017.

\bibitem{Huber_2016_Implicit}
O.~Huber, V.~Acary, B.~Brogliato, and F.~Plestan, ``Implicit discrete-time
  twisting controller without numerical chattering: Analysis and experimental
  results,'' \emph{Control Engineering Practice}, vol.~46, pp. 129 -- 141,
  2016.

\bibitem{Brogliato_2018_STA}
B.~Brogliato, A.~Polyakov, and D.~Efimov, ``The implicit discretization of the
  super-twisting sliding-mode control algorithm,'' in \emph{Proceedings of The
  15th International Workshop on Variable Structure System (VSS)}, Graz,
  Austria, 2018.

\bibitem{Kikuuwe_2010_PSMC}
R.~Kikuuwe, S.~Yasukouchi, H.~Fujimoto, and M.~Yamamoto, ``{P}roxy-based
  sliding mode control: a safer extension of {PID} position control,''
  \emph{{IEEE} Trans. on Robotics}, vol.~26, no.~4, pp. 670--683, 2010.

\bibitem{Jin_2014_AR}
S.~{Jin}, R.~{Kikuuwe}, and M.~{Yamamoto}, ``Improving velocity feedback for
  position control by using a discrete-time sliding mode filtering with
  adaptive windowing,'' \emph{{Advanced Robotics}}, vol.~28, no.~14, pp.
  943--953, 2014.

\end{thebibliography}

\end{document}